\documentclass[twocolumn]{revtex4}

\usepackage{graphicx}
\usepackage{dcolumn}
\usepackage{amsmath}
\usepackage{amssymb}

\newcommand{\dotprod}{{\scriptscriptstyle \stackrel{\bullet}{{}}}}

\begin{document}
\title{The stochastic dynamics of micron and nanoscale elastic cantilevers in fluid: fluctuations from dissipation}
\author{M.R. Paul}
\email{mrp@vt.edu}
\author{M.T. Clark}
\affiliation{Department of Mechanical Engineering, Virginia
Polytechnic and State University, Blacksburg, Virginia 24061}
\author{M.C. Cross}
\affiliation{Department of Physics, California Institute of
Technology, Pasadena, California 91101}

\begin{abstract}The stochastic dynamics of micron and nanoscale cantilevers
immersed in a viscous fluid are quantified. Analytical results are
presented for long slender cantilevers driven by Brownian noise.
The spectral density of the noise force is not assumed to be white
and the frequency dependence is determined from the
fluctuation-dissipation theorem. The analytical results are shown
to be useful for the micron scale cantilevers that are commonly
used in atomic force microscopy. A general thermodynamic approach
is developed that is valid for cantilevers of arbitrary geometry
as well as for arrays of multiple cantilevers whose stochastic
motion is coupled through the fluid. It is shown that the
fluctuation-dissipation theorem permits the calculation of
stochastic quantities via straightforward deterministic methods.
The thermodynamic approach is used with deterministic finite
element numerical simulations to quantify the autocorrelation and
noise spectrum of cantilever fluctuations for a single micron
scale cantilever and the cross-correlations and noise spectra of
fluctuations for an array of two experimentally motivated
nanoscale cantilevers as a function of cantilever separation. The
results are used to quantify the noise reduction possible using
correlated measurements with two closely spaced nanoscale
cantilevers.
\end{abstract}

\maketitle

\section{Introduction}

The dynamics of micron and nanoscale cantilevers are important to
a wide variety of technologies. For example, the invention of the
atomic force microscope (AFM)~\cite{binnig:1986}, which relies
upon the dynamics of a cantilever a few hundred microns in length,
has revolutionized surface science paving the way for direct
measurements of intermolecular forces and topographical mappings
with atomic precision for a broad array of materials including
semiconductors, polymers, carbon nanotubes, and biological
cells~\cite{martin:1987,albrecht:1991,radmacher:1992,zhong:1993,hansma:1994,garcia:2002}
(see~\cite{giessibl:2003,jalili:2004} for current reviews). In
conventional dynamic atomic force microscopy the cantilever is
used to measure the force interactions between the cantilever tip
and sample. Cantilevers smaller than conventional AFM have also
been used to unfold single protein molecules with improved force
and time resolution~\cite{viani:1999}. It has also been proposed
to exploit the inherent thermal motion of small cantilevers to
make dynamic measurements of single
molecules~\cite{arlett:2003,paul:2004:bionems}. A passive undriven
cantilever placed in a viscous fluid will exhibit stochastic
oscillations caused by the thermal bombardment of fluid molecules
by Brownian motion. In fact, measuring the thermal spectra of the
cantilever is a commonly used AFM calibration
technique~\cite{albrecht:1987,sader:1999}.

In all of these applications the ultimate force sensitivity of a
particular measurement is limited by the inherent thermal noise of
the experimental system. There are at least two ways to improve
upon this limitation: (i) make the cantilevers
smaller~\cite{walters:1996,viani:1999}, (ii) use correlated
measurements of multiple
cantilevers~\cite{roukes:2000:1,paul:2004:bionems}.

Uniformly decreasing the dimensions of a cantilever results in the
favorable combination of decreasing the cantilever's equivalent
spring constant while increasing its resonant frequency yielding
improved force sensitivity and time resolution. By measuring the
cross-correlations between two cantilevers in fluid the
independent fluctuations of the two do not contribute, leaving
only the smaller correlated fluctuations due to coupling through
the fluid. This type of approach has been used to measure
femtonewton forces on millisecond time scales between two micron
scale beads placed in separate optical traps (an improvement of a
hundredfold from prior measurements)~\cite{meiners:1999}. The
ability to significantly increase the force resolution by making
correlated measurements has yet to be exploited for micron and
nanoscale cantilevers. The magnitude of the fluid coupled noise
will depend upon the spacing and exact geometries of the
cantilevers. Combining (i) and (ii) and measuring the correlations
of multiple nanoscale cantilevers offer the potential for
experimental measurements with unprecedented force and time
resolution~\cite{roukes:2000:1,arlett:2005,paul:2004:bionems}.

As experimental measurement continues to push toward the
stochastic limit it is important that we build a physical
understanding of the stochastic dynamics of micron and nanoscale
cantilevers for the precise conditions of experiment including
complex cantilever geometries~\cite{albrecht:1987,arlett:2005},
the effects of nearby
walls~\cite{green:2005,clarke:2005,clarke:2006,ma:2000}, and the
fluid-coupled dynamics of multiple cantilevers in an array
configuration~\cite{paul:2004:bionems}.

Although micron and nanoscale cantilevers exhibit stochastic
motion due to the thermal motion of matter the elastic structures
are still large compared to individual fluid molecules and the
equations of continuum mechanics remain valid. In what follows we
are concerned with situations where the Knudsen number (the ratio
of the mean free path of the fluid particles to the width of a
cantilever) remains sufficiently small so that this statement
remains true. This means that the fluid is described by the usual
Navier-Stokes equations with no-slip and stress continuity
boundary conditions at the solid surfaces.

The Navier-Stokes equations governing the motion of an
incompressible fluid, and written in nondimensional form, are
\begin{eqnarray}
R_\omega~\frac{\partial \vec{u}}{\partial t} + R_u~\vec{u}
\dotprod
\vec{\nabla} \vec{u} &=& -\vec{\nabla} p + \nabla^2 \vec{u}, \label{eq:ns} \\
\vec{\nabla} \dotprod \vec{u} &=& 0, \label{eq:mass}
\end{eqnarray}
where $\vec{u}$ is the fluid velocity, $p$ is the pressure, and
$t$ is time. There are two inertial terms on the left hand side of
Eq.~(\ref{eq:ns}) multiplied by the nondimensional parameters
$R_{\omega}$ and $R_u$. The Strouhal number $R_{\omega} = L^2/\nu
T$ plays the role of a frequency based Reynolds number expressing
the ratio between local inertia forces and viscous forces where
$L$ and $T$ are characteristic length and time scales,
respectively. The velocity based Reynolds number $R_u=UL/\nu$
expresses the ratio between convective inertial forces and viscous
forces.

Micron and nanoscale cantilevers are characterized by high
oscillation frequencies and small oscillation amplitudes. In this
case $R_u \ll 1$ so that the nonlinear convective inertial term
$\vec{u} \dotprod \vec{\nabla} \vec{u}$ is negligible and the
equations become linear. However, the Strouhal number,
\begin{equation}
 R_\omega = \frac{\omega w^2} {4 \nu},
 \label{eq:R_omega}
\end{equation}
(using the half-width $w/2$ as the appropriate length scale) is
often not negligible. As a result, the local inertia term must be
kept in Eq.~(\ref{eq:ns}) making the analysis more difficult. In
addition, experimentally motivated cantilevers are often of
complex geometry, near surfaces, or in an array configuration with
multiple cantilevers in close proximity. These difficulties have
led to the development of a thermodynamic approach to calculate
the stochastic dynamics of micron and submicron scale cantilevers
for the precise conditions of experiment (discussed
below)~\cite{paul:2004:bionems}. In the following it is assumed
that $R_u$ is negligible, and we leave off the subscript denoting
$R_\omega$ by $R$.

The fluid equations are coupled with the governing equations of
elasticity,
\begin{equation}
\rho_c \frac{\partial^2 \vec{w}}{\partial t^2} = \vec{\nabla}
\dotprod \mathbf{\sigma}, \label{eq:beam}
\end{equation}
where $\sigma$ is the stress tensor, $\rho_c$ is the cantilever
density, and $\vec{w}(x,y,z,t)$ is the cantilever deflection.
Equations~(\ref{eq:ns}),(\ref{eq:mass}), and (\ref{eq:beam})
represent the governing deterministic continuum equations. If one
considers lengths scales on the order of a few atomic lengths
these equations would include stochastic
terms~\cite{landau:1959:fluids}. This represents a difficult
fluid-solid interaction problem governing the stochastic
cantilever dynamics. A brute force molecular dynamics approach
which would resolve the stochastic motion of all of the fluid and
solid molecules is computationally prohibitive. However, the
system is always near equilibrium permitting a much more
accessible thermodynamic based solution strategy.

The paper is organized as follows. A very general thermodynamic
approach, valid for complex geometries as well as for arrays of
closely spaced cantilevers, is discussed in
Section~\ref{section:thermodynamic_approach}. Analytical theory
based upon the thermodynamic approach, valid for long slender
cantilevers, is discussed in
Section~\ref{section:analytical_theory}. In
Section~\ref{section:micron_scale_beam} a micron scale cantilever,
similar to an atomic force microscope, is explored and in
Section~\ref{section:nanoscale_cantileves} the stochastic dynamics
of an array of experimentally motivated nanoscale cantilevers are
quantified using the thermodynamic approach with deterministic
finite element numerical simulations to generate results for the
precise conditions of experiment. A general discussion of the
approach can be found in Ref.~\cite{paul:2004:bionems}.
\begin{figure}[tbh]
\begin{center}
\includegraphics[width=3.0in]{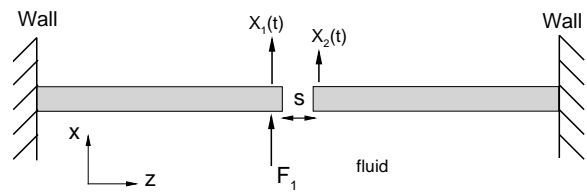}
\end{center}
\caption{Schematic illustrating an array of two opposing
cantilever beams separated by a distance $s$. In the thermodynamic
approach of Section~\ref{section:thermodynamic_approach} a step
force $F_1$ is removed from the cantilever on the left and the
deterministic cantilever deflections $X_1(t)$ and $X_2(t)$ are
calculated.} \label{fig:adjacent}
\end{figure}

\section{Thermodynamic approach}
\label{section:thermodynamic_approach}

The thermodynamic approach discussed here is based upon the
fluctuation-dissipation theorem which states that for equilibrium
systems the manner in which the system returns from a linear
macroscopic perturbation is related to time correlations of
equilibrium microscopic
fluctuations~\cite{callen:1951,callen:1952,chandler:1987}.
Underlying this statement is the fact that equilibrium
fluctuations and the dissipation of the system responding to a
macroscopic perturbation are governed by the same physics. For the
case of miniature elastic objects in a fluid the dissipation is
dominated by the fluid viscosity and the fluctuations by the
motion due to the bombardment by the fluid molecules. In what
follows, we assume that all of the dissipation comes from the
fluid and that elastic dissipation in the cantilever is
negligible. This assumption, however, is not required and other
sources of dissipation could be included if desired. The essence
of our approach is to calculate the dissipation
\textit{deterministically} and to use the fluctuation-dissipation
theorem to determine the cantilever's \text{stochastic} dynamics.
The approach is exact and the only assumptions that have been made
are classical dynamics and linear perturbations from equilibrium.
The deterministic calculation of the dissipation can come from
analytical theory, simplified models, or from detailed numerical
simulations. The major benefit of this approach is that the
deterministic calculations are straightforward, not
computationally prohibitive, and methods of calculation are
sophisticated and readily available.

We will introduce the use of the thermodynamic approach for the
case of two opposing cantilevers as shown in
Fig.~\ref{fig:adjacent}. Consider one dynamical variable to be the
stochastic displacement of the cantilever on the left $x_1(t)$
where $x_1(t)$ is a function of the microscopic phase space
variables consisting of $3N$ coordinates, $r^{3N}$, and conjugate
momenta, $p^{3N}$, of the system and $N$ is the number of
particles in the system. The statistical treatment that follows
determines the ensemble average of the cantilever deflections
$x_1(t)$ over all experimental possibilities. For the undisturbed
equilibrium system the ensemble average, and correlations, are
denoted as $\left< \right>$.

We now take the system to a prescribed macroscopic excursion from
equilibrium and observe how the system returns to equilibrium. The
macroscopic cantilever deflection as it returns to equilibrium
from a prepared initial condition will be denoted as $X_1(t)$. For
clarity of presentation, lower case variables are reserved for
stochastic quantities and upper case variables are for
deterministic quantities. The connection between the stochastic
and deterministic quantities is particularly straightforward when
the excursion from equilibrium is achieved through the application
of a step force some time in the distant past that is removed at
time $t=0$, i.e. the force given by,
\begin{equation}
f(t)=\left\{
\begin{array}
[c]{cc}%
F_1 & \text{for }t \le 0\\
0 & \text{for }t > 0.
\end{array}
\right. \label{eq:step_force}
\end{equation}
The applied force should be conjugate to the variable for which
the fluctuations are to be calculated. For the cantilever system
under consideration here, to determine the fluctuations of the tip
displacements, the force is applied to the tip of the left
cantilever. The step force couples with the cantilever deflection
$x_1(t)$ and the full Hamiltonian of the system $H$ is given by,
\begin{equation}
H=\left\{
\begin{array}
[c]{cc}%
H_0 + \Delta H & \text{for }t \le 0\\
H_0 & \text{for }t > 0
\end{array}
\right. \label{eq:hamiltonian}
\end{equation}
where $\Delta H = -F_1 x_1(t) = -F_1 x_1(0)$ and $H_0$ is the
unperturbed Hamiltonian. For small perturbations, $F_1$, the
quantity $\Delta H$ will also be small, and the results are
greatly simplified. The nonequilibrium ensemble average $X_1(t)$
is given by,
\begin{equation}
X_1(t) = \frac{\int dr^{3N} dp^{3N} x_1(t) e^{-\beta(H_0+ \Delta
H)}} {\int dr^{3N} dp^{3N} e^{-\beta(H_0 + \Delta H)}},
\label{eq:noneq_ave}
\end{equation}
where $\beta = (k_B T)^{-1}$, $k_B$ is Boltzmann's constant, and
$T$ is the absolute temperature. In Eq.~(\ref{eq:noneq_ave}) the
notation $x_1(t)$ is used to represent the value of $x_1$
evaluated at the phase space coordinates that have evolved from
the values $\vec{r}$ and $\vec{p}$ at time
$t=0$~\cite{chandler:1987}. In the limit of linear perturbations,
$\Delta H \ll 1$, this simplifies to,
\begin{equation}
X_1(t) = \left< x_1 \right> + \beta \left< \Delta H \right> \left<
x_1 \right> - \beta \left< x_1 \Delta H \right> +
\mathcal{O}(\Delta H^2),
\end{equation}
where an equilibrium ensemble average is given by,
\begin{equation}
\left< x_1(t) \right> = \frac{\int dr^{3N} dp^{3N} x_1(t)
e^{-\beta H_0}} {\int dr^{3N} dp^{3N} e^{-\beta H_0}}.
\end{equation}
If we now assume that $x_1$ and $X_1$ have the equilibrium average
$\left< x_1 \right>$ subtracted and recall that $\Delta H = -F_1
x_1(0)$, we have our desired result,
\begin{equation}
X_1(t) = \beta F_1 \left< x_1(t) x_1(0) \right>
\label{eq:fdt_displacements}
\end{equation}
which can be rearranged to yield,
\begin{equation}
\left< x_1(0) x_1(t) \right>  = k_B T\frac{F_1}{X_1(t)}.
\label{eq:auto}
\end{equation}
The analysis is similar for the cross correlations of the
deflections for two cantilevers in an array,
\begin{equation}
\left< x_1(0) x_2(t) \right>  = \frac{k_B T}{F_1} X_2(t),
\label{eq:cross}
\end{equation}
where $X_2$ is the displacement of tip 2 arising from the step
force $F_1$ applied to tip 1.

It is interesting to highlight that the cantilever motions are
indeed correlated as indicated by Eq.~(\ref{eq:cross}). The
correlated fluctuations appear contrary to the naive idea that
random molecular impacts upon the individual cantilevers should
only lead to uncorrelated motion. Additionally, it should be
emphasized that the correlated motion of two (or more) cantilevers
is no more difficult a calculation than determining the
autocorrelation of a single cantilever.

On the left hand side of Eqs.~(\ref{eq:auto}) and~(\ref{eq:cross})
are stochastic quantities and on the right hand side are
deterministic quantities. This permits a significant reduction in
effort in the calculations of the auto- and cross-correlations of
the cantilever deflections. The macroscopic deflections $X_1(t)$
and $X_2(t)$ can be found from deterministic theory, model
equations, or numerical simulations and the stochastic quantities
are found simply through application of Eqs.~(\ref{eq:auto})
and~(\ref{eq:cross}). The deterministic problem then reduces to
the fluid-solid interaction problem given by the removal of a step
force on a cantilever. For the step force
Eq.~(\ref{eq:step_force}), for $t<0$, $X_1(t)$ has a finite
deflection. After the removal of the step force the cantilever
returns to its equilibrium position of $X_1 = 0$. The precise
manner in which the cantilever returns to equilibrium is dominated
by the dissipation in the fluid. An adjacent cantilever will
exhibit time dependent deflections given by $X_2(t)$. In this case
both the initial and final equilibrium deflections are $X_2=0$.
The motion of the second cantilever is a result of the fluid
motion caused by the first cantilever as it returns to
equilibrium.

The spectral properties of the auto- and cross-correlations are
determined by taking the cosine Fourier transform of
Eqs.~(\ref{eq:auto}) and~(\ref{eq:cross}) with appropriate factors
(see Eq.~(\ref{eq:spectral_density_definition})). This yields the
noise spectra, $G_{11}(\omega)$ and $G_{12}(\omega)$, given by,
\begin{eqnarray}
G_{11}(\omega) &=& 4 \int^\infty_0 \left< x_1(0) x_1(t) \right>
\cos(\omega t) dt, \label{eq:noise1}\\
G_{12}(\omega) &=& 4 \int^\infty_0 \left< x_1(0) x_2(t) \right>
\cos(\omega t) dt, \label{eq:noise2}
\end{eqnarray}
where $\omega$ is the angular frequency. The noise spectra are
precisely the measured quantity in experiment. In the above
expression we have defined the spectral density of the random
process $y(t)$ to be,
\begin{equation}
G_y(\omega) = \lim_{T \rightarrow \infty} \frac{1}{\pi T} \left|
\int_{-T/2}^{T/2} [y(t) - \bar{y}] e^{i \omega t} dt \right|^2,
\label{eq:spectral_density_definition}
\end{equation}
where $\bar{y}$ is the average of $y(t)$ over time $T$.

In summary, the explicit steps necessary to determine the
stochastic dynamics of a cantilever array are:
\begin{enumerate}
\item Perform the \textit{deterministic} calculation. Given some
arrangement of cantilevers in a fluid choose one cantilever and
apply a step force some time in the distant past. This will
eventually result in a finite steady deflection of this
cantilever, whereas the other cantilevers will have zero
deflection.

\item Remove the step force and calculate the deflections of the
cantilevers as a function of time as they return to equilibrium,
for two cantilevers this is $X_1(t)$ and $X_2(t)$ (see for example
Fig.~\ref{fig:beam_fit} for the case of one cantilever). This
deterministic calculation can be done analytically or numerically
depending upon the complexity of the particular situation.

\item Calculate the \textit{stochastic} quantities. Using $X_1(t)$
and $X_2(t)$ calculate the auto- and cross-correlations of the
fluctuations from Eqs.~(\ref{eq:auto}) and~(\ref{eq:cross}). In
essence, the cantilever deflections are merely rescaled to produce
the stochastic response (this is illustrated by the two ordinate
axes in Fig.~\ref{fig:beam_fit}).

\item Use the auto- and cross-correlations to calculate the noise
spectra given by Eqs.~(\ref{eq:noise1}) and~(\ref{eq:noise2}).
\end{enumerate}

\section{Analytical theory based on an oscillating infinite
cylinder}\label{section:analytical_theory}

\begin{figure}[h]
\begin{center}
\includegraphics[width=3.0in]{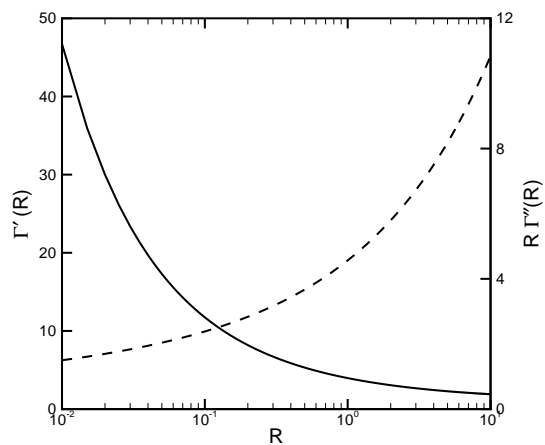}
\end{center}
\caption{The frequency dependence of the fluid loaded mass and
fluid damping for an oscillating cantilever in a viscous fluid.
(solid line) $\Gamma'(R)$ illustrates the frequency dependence of
the fluid loaded mass. (dashed line) $R \Gamma''(R)$ illustrates
the frequency dependence of the fluid damping.}
\label{fig:mass_damping}
\end{figure}

The equipartition theorem permits the calculation of the mean
square mode displacements of micron and nanoscale cantilevers in
vacuum~\cite{butt:1995}. Sader~\cite{sader:1998} further developed
this approach to include the damping forces of a surrounding
viscous fluid and solved the equations of elasticity in the thin
beam limit coupled together with the Navier-Stokes equations. He
used the approximation that the effect of the fluid on each
element of cantilever is the same for an element of a long and
slender cantilever moving with the same speed. In the limit of an
infinitely long cylinder the flow field around the cantilever can
be assumed to be the flow field around an infinite two-dimensional
cylinder to a good approximation~\cite{sader:1998,tuck:1969}. In
addition, because the forces on each element of cantilever due to
the fluid depend only on the velocity of that element, and not
otherwise on its position in the cantilever, the mode structure of
the damped cantilever is unchanged from the undamped limit. We
follow this approach here and use the infinite cylinder
approximation for the flow field. However we do not assume that
the fluid noise is white as was done in~\cite{sader:1998} but use
the correct spectral density of the Brownian force given by
$G_F(\omega)$ in Eq.~(\ref{eq:fdt}). The frequency dependence of
the Brownian noise $G_F(\omega)$ is determined by the fluid
damping $\gamma_f(\omega)$ (illustrated in
Fig.~\ref{fig:mass_damping} and discussed further below).

To connect with the previous work we note that the susceptibility $\chi(t)$
connecting the tip displacement to an applied force
is given by the cantilever response to a unit impulse of force,
which is the derivative of the step force response given by
Eq.~(\ref{eq:auto}) so that
\begin{equation}
\chi(t) = - \beta \frac{d}{dt} \left< x_1(0) x_1(t) \right>.
\label{eq:kubo}
\end{equation}
The Fourier transform of $\chi(t)$ is the response to a
deterministic sinusoidal force and the imaginary part
$\hat{\chi}''(\omega) = \text{Im} \left\{\hat{\chi}(\omega)
\right\}$, where Im\{\} indicates the imaginary component is,
\begin{equation}
\hat{\chi}''(\omega) = -\beta \int_0^{\infty} \frac{d}{dt} \left<
x_1(0) x_1(t) \right> \sin (\omega t ) dt, \label{eq:dissipation}
\end{equation}
where we have adopted the Fourier transform convention given by,
\begin{eqnarray}
\hat{x}(\omega) &=& \int_{-\infty}^{\infty} x(t) e^{i \omega t}
dt \\
x(t) &=& \frac{1}{2 \pi} \int_{-\infty}^{\infty} \hat{x}(\omega)
e^{- i \omega t} d \omega,
\end{eqnarray}
and $i = \sqrt{-1}$. Integrating Eq.~(\ref{eq:dissipation}) by
parts and using the definition of the noise spectrum
Eq.~(\ref{eq:noise1}) yields,
\begin{equation}
G_{11}(\omega) = \frac{4 k_B T}{\omega} \hat{\chi}''(\omega).
\label{eq:gxf_susceptibility}
\end{equation}
The susceptibility $\chi(t)$ can be determined by solving for the
deterministic cantilever response to a force impulse.

In the following we present results only for the fundamental mode;
however higher modes can be included if desired. The equation
governing the cantilever motion is then,
\begin{equation}
m_e \ddot{X_i} + k X_i = F_f + F_i,
\end{equation}
where the amplitude of the mode displacement is characterized by
the tip displacement $X_i$, $m_e$ is the effective mass of the
cantilever in vacuum, $F_{f}$ is the force acting on the
cantilever due to the fluid, and $F_i = \delta(t)$ is the force
impulse ($\delta(t)$ is the Dirac delta). The effective mass of
the cantilever $m_e$ is chosen to yield the same kinetic energy as
in the cantilever mode, and is related to the cantilever mass by,
\begin{equation}
 m_e= \alpha m_c,
\end{equation}
where $m_c$ is the actual cantilever mass and for the fundamental
mode of oscillation for a beam $\alpha = 0.243$. Taking the
Fourier transform of this equation yields,
\begin{equation}
 \left( -m_e \omega^2 + k \right)\hat{X_i} = \hat{F}_{f} + 1.
 \label{eq:eom_fourier_deterministic}
\end{equation}
The force from the fluid can be written in the form,
\begin{equation}
\hat{F}_{f} = m_{cyl,e} \omega^2 \Gamma(\omega) \hat{x},
\end{equation}
where
\begin{equation}
  m_{cyl,e} = \alpha m_{cyl} = \alpha \rho_f \left( \frac{\pi}{4} w^2 L
  \right).
\end{equation}
Here $m_{cyl}$ is the effective mass of a fluid cylinder of radius
$w/2$ where $\rho_f$ is the fluid density, $w$ is the cantilever
width, and $L$ is the cantilever length. Again the prefactor
$\alpha=0.243$ is to take into account the variation of the fluid
force along the cantilever due to the varying velocity given by
the mode structure. The fluid loading and damping are captured by
the hydrodynamic function $\Gamma(\omega)$ given by,
\begin{equation}
\Gamma(\omega) = 1 + \frac{4i
K_1(-i\sqrt{iR})}{\sqrt{iR}K_0(-i\sqrt{iR})},
\end{equation}
where $K_1$ and $K_0$ are Bessel functions~\cite{rosenhead:1963}.
In this definition the frequency dependence on the right hand side
appears through the frequency dependent Strouhal number $R$.

The cantilever is loaded by the fluid which can be characterized
by an effective mass, $m_f$, larger than $m_e$ that takes into
account the fluid mass that is also being moved. The fluid also
damps the motion of the cantilever which can be expressed as an
effective damping $\gamma_f$. Relations for $m_f$ and $\gamma_f$
can be found by expanding $\Gamma(\omega)$ into its real and
imaginary parts in Eq.~(\ref{eq:eom_fourier_deterministic}) and
rearranging such that,
\begin{equation}
-m_f(\omega)\omega^2\hat{X_i} - i \omega \gamma_f(\omega)
\hat{X_i} + k \hat{X_i} = 1 \label{eq:fourier_loaded}
\end{equation}
to give,
\begin{equation}
 m_f = \alpha m_c\left( 1 + T_0 \Gamma' \right)
 \label{eq:fluid_loaded_mass}
\end{equation}
and,
\begin{equation}
 \gamma_f = \alpha m_{cyl} \omega \Gamma'',
 \label{eq:damping}
\end{equation}
where $\Gamma'$ and $\Gamma''$ are the real and imaginary parts of
$\Gamma$, respectively. $T_0$ is the mass loading parameter, which
is the ratio of the mass of a cylinder of fluid with radius $w/2$
to the actual mass of the cantilever, and is given by,
\begin{equation}
T_0 = \frac{m_{cyl}}{m_c} = \frac{\pi}{4} \frac{\rho_f w}{\rho_c
h}, \label{eq:T0}
\end{equation}
where $\rho_c$ is the density of the cantilever and $h$ is the
cantilever thickness.

It is evident from Eqs.~(\ref{eq:fluid_loaded_mass})
and~(\ref{eq:damping}) that both the fluid loaded mass of the
cantilever and the fluidic damping are functions of frequency. The
ratio of the mass of the fluid loaded cantilever to the effective
mass of the cantilever in vacuum, $m_e$, as a function of
frequency is given by $m_f/m_e = 1 + T_0 \Gamma'(R)$. The factor
$T_0$ is a constant for any particular choice of cantilever and
fluid combination. The frequency dependence of the added mass is
then given by $\Gamma'(R)$ and is illustrated by the solid line in
Fig.~\ref{fig:mass_damping} and uses the left ordinate axis. The
added mass increases rapidly as the frequency is decreased. Over
the range of four orders of magnitude in frequency the added mass
is seen to change by a factor of approximately 25. The
frequency dependence of the fluid damping is given by $\omega
\Gamma''(\omega)$ and is shown by the dashed line in
Fig.~\ref{fig:mass_damping} using the right ordinate axis. There
is a weaker frequency dependence for the fluid damping when
compared to the mass loading. The fluid damping decreases as the
frequency decreases and over 4 orders of magnitude in frequency
the damping changes by a factor of approximately 7.

Solving for the cantilever response $\hat{X}_i(\omega)$ and taking
the imaginary part yields,
\begin{equation}
\hat{\chi}'' = \frac{\omega \gamma_f}{(-m_f \omega^2 + k^2)^2 -
(\omega \gamma_f)^2}.
\end{equation}
Using this result in Eq.~(\ref{eq:gxf_susceptibility}) and
rearranging gives the desired result for the spectral density of
the stochastic fluctuations in cantilever displacement,
\begin{eqnarray}
\lefteqn{G_{11}(\omega) =  \frac{4k_BT}{k}\frac{1}{\omega_0} \dotprod}  \label{eq:gxf_noise} \\
& & \frac{T_0 \tilde{\omega} \Gamma_i(R_0 \tilde{\omega})}{\left[
\left(1 - \tilde{\omega}^2 \left( 1 + T_0 \Gamma_r(R_0
\tilde{\omega}) \right)\right)^2 + \left( \tilde{\omega}^2 T_0
\Gamma_i(R_0 \tilde{\omega}) \right)^2 \right] \nonumber},
\end{eqnarray}
where $\tilde{\omega}=\omega/\omega_o$ is a nondimensional reduced
frequency, and $R_0$ is the Strouhal number evaluated at the
resonant frequency in vacuum given by,
\begin{equation}
  R_0 = \frac{\rho_f \omega_0 w^2}{4 \eta}. \label{eq:R0}
\end{equation}
We emphasize that Eq.~(\ref{eq:gxf_noise}) is \emph{not} the same as in previous
work~\cite{sader:1998} which failed to include the frequency dependent of the
Brownian force.

To understand the difference from the previous work we now connect Eq.~(\ref{eq:fdt_displacements}) with
the calculation in terms of a fluctuating force with spectral
density $G_F(\omega)$. The fluctuating
displacement can be described in terms of the response to this force through the
susceptibility
\begin{equation}
G_{11}(\omega) = |\hat{\chi}(\omega)|^2 G_F(\omega).
\end{equation}
Inserting Eq.~(\ref{eq:gxf_susceptibility}) into this
expression yields,
\begin{equation}
G_F(\omega) = \frac{4 k_B T}{\omega} \text{Im}\left\{ -
\frac{1}{\hat{\chi}(\omega)} \right\}.
\label{eq:gf_susceptibility}
\end{equation}
It is useful at this point to introduce the impedence $Z(t) = F/v$
where $v$ is the velocity of the cantilever tip $\left<
\dot{x}_1(t) \right>$, it is then straight forward to show,
\begin{equation}
\hat{Z}(\omega) = -\frac{1}{i \omega \hat{\chi}(\omega)}.
\end{equation}
Inserting this into Eq.~(\ref{eq:gf_susceptibility}) and
rearranging yields the desired expression of the fluctuation
dissipation theorem,
\begin{equation}
G_F(\omega) = 4 k_B T \gamma_f(\omega) \label{eq:fdt}
\end{equation}
where $\gamma_f(\omega) = \text{Re} \left\{
\hat{Z}(\omega)\right\}$ is the resistance or dissipation and
Re\{\} indicates the real part. For the case of oscillating
cantilevers in fluid $\gamma_f$ is the effective fluid damping. It
is evident from the frequency dependence of the damping that the
fluctuating force described by Eq.~(\ref{eq:fdt}) is not white
noise as considered previously~\cite{sader:1998} (see
Fig.~\ref{fig:mass_damping} for the variation of $\gamma_f$ with
frequency).

The spectral density of fluctuations in cantilever displacement
$G_{11}(\omega)$ can also be determined by directly solving the
governing stochastic equation of motion,
\begin{equation}
m_e\ddot{x_1} + kx_1 = F_{f} + F_B,
\end{equation}
where $F_B$ is the random force due to Brownian motion. Taking the
Fourier transform of this equation yields,
\begin{equation}
 \left( -m_e \omega^2 + k \right)\hat{x} = \hat{F}_{f} + \hat{F}_{B}.
 \label{eq:eom_fourier}
\end{equation}
Solving for the magnitude of the cantilever response and using
Eq.~(\ref{eq:fdt}) for the spectral density of the Brownian force
again yields Eq.~(\ref{eq:gxf_noise}). Note that integrating this result
automatically leads to the equipartition result,
\begin{equation}
 \frac{1}{2 \pi} \int^{\infty}_{0} G_{11}(\omega) d\omega = \frac{k_B T}{k}.
 \label{eq:equipartition}
\end{equation}
We would like to point out that it is not possible to start with
the equipartition result given by Eq.~(\ref{eq:equipartition}) and
to then calculate the spectral properties of the fluctuations, one
must first start with the fluctuation-dissipation result given by
Eq.~(\ref{eq:fdt}) as done here.

The noise spectrum of the cantilever in fluid may be used to
calibrate an AFM. A simple way to do this is to extract the
effective spring constant from the the frequency $\omega_f$ giving
the maximum of the noise intensity. In previous
work~\cite{sader:1998} the fluctuations of the cantilever have
been estimated without the frequency dependence of the numerator
in Eq.~(\ref{eq:gxf_noise}), leading to some inaccuracy in the
estimate of the spring constant. Once $\omega_f$ is known, the
quality factor $Q$ can be estimated from,
\begin{equation}
 Q \approx \frac{\omega m_f}{\gamma_f} = \frac{\frac{1}{T_0} +
\Gamma_r(R)}{\Gamma_i(R)}, \label{eq:Qapprox}
\end{equation}
where Eqs.~(\ref{eq:fluid_loaded_mass}) and~(\ref{eq:damping})
have been used for the mass and damping. This expression for $Q$ is only
approximate again because the frequency response of the cantilever
is not precisely like that of simple harmonic oscillator, since
both $m_f$ and $\gamma_f$ are frequency dependent.

\begin{figure}[h]
\begin{center}
\includegraphics[width=3.25in]{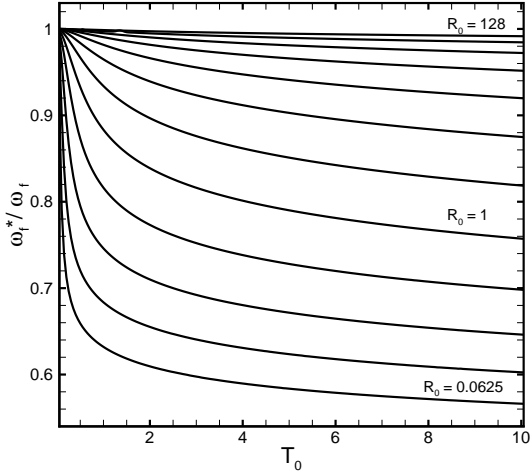}
\end{center}
\caption{The error in predicting the frequency at the maximum of
the noise spectrum $\omega_f^*/\omega_f$. $\omega_f^*$ is the
predicted frequency at the maximum of the noise spectrum when the
frequency dependence of the Brownian force is neglected as done in
previous work~\cite{sader:1998} and $\omega_f$ is the correct
value given by the results of
Section~\ref{section:analytical_theory}. Results are presented as
a function of mass loading $T_0$ and the Strouhal number $R_0$
based upon the cantilever's resonant frequency in vacuum. The
curves are for $R_0=128,64,32,16,4,1,0.25,0.0625$ where
$R_0=0.0625,1,128$ are labelled and the remaining curves are in
sequential order.} \label{fig:mass_loading_noise_compare}
\end{figure}

The error incurred by neglecting the frequency dependence in the
numerator in Eq.~(\ref{eq:gxf_noise}) in fitting the spectrum of a
Brownian driven cantilever is illustrated in
Fig.~\ref{fig:mass_loading_noise_compare} as a function of $R_0$
and $T_0$. In Fig.~\ref{fig:mass_loading_noise_compare}
$\omega_f^*$ is the frequency at the maximum of the noise spectrum
when the frequency dependence has not been included (specifically,
in Eq.~(\ref{eq:gxf_driven}) $G_F(\omega)$ has been assumed
constant) and $\omega_f$ is the correct value of the frequency
using the fluctuation-dissipation theorem as discussed in
Section~\ref{section:analytical_theory}. The results indicate that
as the Strouhal number increases the error in this assumption
becomes smaller. In particular, when the frequency dependence is
neglected the approximate theory always underpredicts $\omega_f$.
This can be understood in light of Fig.~\ref{fig:mass_damping}
where it is clear that the magnitude of the damping decreases as
the frequency of oscillation is reduced. When this decrease is not
included the predicted value of $\omega_f$ will be unnecessarily
reduced. Although the fluid induced damping slowly decreases as
the oscillation frequency is reduced it is important to note that
the fluid loaded mass increases rapidly and it is this interplay
which results in the small $Q$ associated with oscillating micron
and nanoscale cantilevers.

\begin{figure}[h]
\begin{center}
\includegraphics[width=3.25in]{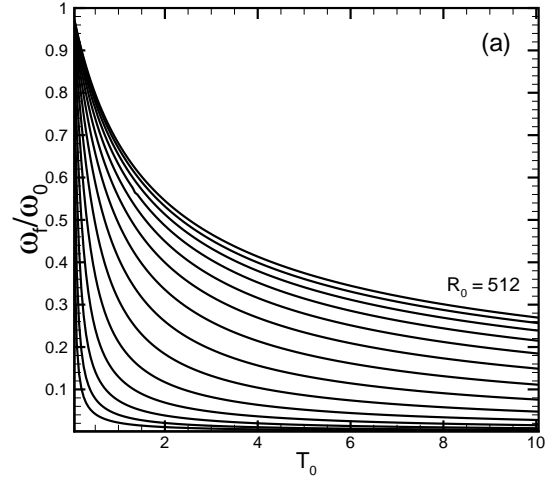}
\includegraphics[width=3.25in]{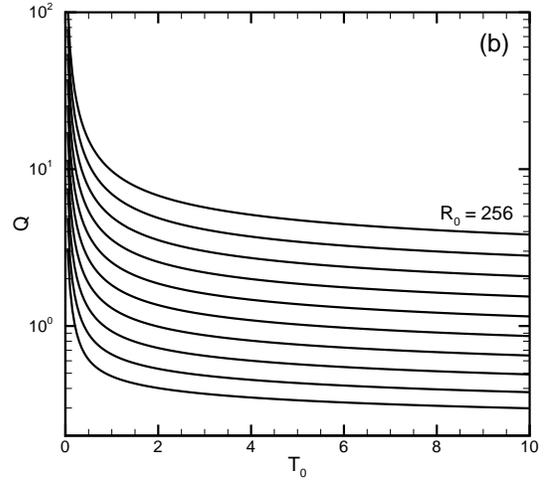}
\end{center}
\caption{Panel~(a) the reduced frequency $\tilde{\omega}$ of
oscillation for a cantilever undergoing stochastic oscillations in
a viscous fluid as a function of mass loading $T_0$ and the
Strouhal number $R_0$ based upon the cantilever's resonant
frequency in vacuum. Panel~(b), the quality factor $Q$ as a
function of $R_0$ and $T_0$ as given by Eq.~(\ref{eq:Qapprox}). In
each panel the largest value of $R_0$ is labelled and each
successive curve represents $R_0/2$, i.e. in panel~(a) the last
curve is for $R_0=0.0625$ and in panel~(b) the last curve is for
$R_0=1/2$.} \label{fig:mass_loading_noise}
\end{figure}

Figure~\ref{fig:mass_loading_noise} summarizes the theoretical
results and allows for easy determination of the stochastic
dynamics of a single cantilever of arbitrary geometry placed in an
arbitrary viscous fluid. Given a particular cantilever and fluid
combination the procedure is:
\begin{enumerate}
\item Determine the cantilever spring constant $k$, resonant
frequency in vacuum, $\omega_0$, and characteristic length scale.
These can be experimental measurements or theoretical
calculations. The characteristic length is a measure of the
characteristic half-width of the cantilever.

\item Determine $R_0$ from Eq.~(\ref{eq:R0}) and $T_0$ from
Eq.~(\ref{eq:T0}).

\item Use $R_0$ and $T_0$ to determine $\omega_f/\omega$ from
Fig.~\ref{fig:mass_loading_noise}(a) and $Q$ from
Fig.~\ref{fig:mass_loading_noise}(b).
\end{enumerate}

In the case that the cantilever is driven externally the response
can be found in a similar manner. If the cantilever is driven by a
force $F_d(t)$ the equation of motion becomes,
\begin{equation}
 m_e\ddot{x} + kx = F_{f} + F_d.
 \label{eq:eom_driven}
\end{equation}
If the driving force is $F_d = A_0 \sin(\omega_d t)$ where $A_0$
is a constant and $\omega_d$ is the driving frequency the
amplitude of the response is given by,
\begin{eqnarray}
\lefteqn{|\hat{x}(\omega_d)|^2 = \left(\frac{A_0 \pi}{k}\right)^2
\dotprod}  \label{eq:gxf_driven} \\ & & \frac{ 1 }{\left[ \left(1
- \tilde{\omega}_d^2 \left( 1 + T_0 \Gamma_r(R_0 \tilde{\omega}_d)
\right)\right)^2 + \left( \tilde{\omega}_d^2 T_0 \Gamma_i(R_0
\tilde{\omega}_d) \right)^2 \right]} \nonumber,
\end{eqnarray}
where $\tilde{\omega}_d = \omega_d/\omega_0$ is the reduced
driving frequency. We emphasize that Eqs.~(\ref{eq:eom_driven})
and~(\ref{eq:gxf_driven}) are for externally driven cantilevers
and neglect Brownian noise. Eqs.~(\ref{eq:gxf_driven})
and~(\ref{eq:gxf_noise}) are similar in that they share a common
denominator, whereas the additional frequency dependence in the
numerator of Eq.~(\ref{eq:gxf_noise}) is from the frequency
dependence of the Brownian noise.

\section{Micron scale beam in fluid}
\label{section:micron_scale_beam} 
\begin{figure}[tbh]
\begin{center}
\includegraphics[width=2.5in]{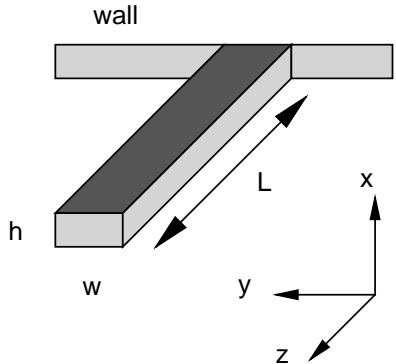}
\end{center}
\caption{Schematic of a simple cantilevered beam of length $L$,
width $w$, and height $h$.} \label{fig:beam}
\end{figure}

The stochastic dynamics of a
micron scale cantilever, similar to what is commonly used in
atomic force microscopy, are now quantified (see
Fig.~\ref{fig:beam}). The beam is chosen from Chon and
Sader~\cite{chon:2000} where both theoretical and experimental
values are present for comparison. The beam properties are
summarized in Table~\ref{table:beam_geometry},
\begin{table}
[h]
\begin{center}
\begin{tabular} [c]{l@{\hspace{0.5cm}}l@{\hspace{0.5cm}}l@{\hspace{0.5cm}}
l@{\hspace{0.5cm}}l@{\hspace{0.5cm}}l@{\hspace{0.5cm}}l}
  $L$      & $w$     & $h$    & $f_0$ & $R_0$ & $T_0$ \\ \hline
  \hline \\
  197$\mu$m & 29$\mu$m & 2$\mu$m & 71.56KHz   & 109.7 & 4.89  \\
\end{tabular}
\end{center}
 \caption{The beam geometry: length $L$, width $w$, thickness
 $h$, resonant frequency in vacuum $f_0$, Strouhal number
 based on the vacuum resonant frequency $R_0$ and the mass loading
 factor $T_0$. The beam is silicon with $E = 1.74 \times 10^{11}$N/m$^2$, $\rho_c =
2320$Kg/m$^3$ and the fluid is water with ($\rho_f = 997$Kg/m$^3$,
$\eta = 8.59 \times 10^{-4}$Kg/ms). In the deterministic numerical
simulations the magnitude of the step force applied is $F_0 =
26$nN.}
 \label{table:beam_geometry}
\end{table}
using classical beam theory~\cite{landau:1959} the equivalent
spring constant $k$ is predicted to be,
\begin{equation}
k = \frac{3EI}{L^3},
\end{equation}
which yields a value of $k=1.3$N/m and the resonant frequency of
the cantilever in vacuum is
\begin{equation}
f_0= \frac{C_1^2}{2 \pi L^2}\sqrt{\frac{EI}{\mu}},
 \label{eq:w0}
\end{equation}
which yields $f_0 = 71.56$KHz where $C_1=1.8751$. Using
Eq.~(\ref{eq:R0}) the Strouhal number in vacuum is $R_0 = 109.7$
and using Eq.~(\ref{eq:T0}) the mass loading factor is $T_0 =
4.89$.
\begin{table}
[h]
\begin{center}
\begin{tabular}[c]{l@{\hspace{0.5cm}}l@{\hspace{0.5cm}}l@{\hspace{0.5cm}}l@{\hspace{0.5cm}}l@{\hspace{0.5cm}}l@{\hspace{0.5cm}}l}
  & $Q$ & $m_f/m_e$ & $\omega_f/\omega_0$ & $\gamma_f$ (kg/s) & $R_f$  \\
  \hline \hline \\
  (1)  & 3.24 & 8.16 & 0.34 & $5.07 \times 10^{-7}$ & 37.3 \\
  (2)  & 2.93 & 8.05 & 0.35 & $6.91 \times 10^{-7}$ & 38.7 \\
\end{tabular}
\end{center}
 \caption{The stochastic dynamics of a single micron scale cantilever in fluid. (1) Results
 based on the analytical predictions of Section~\ref{section:analytical_theory}. The quality factor $Q$
 is calculated from the approximation given by Eq.~(\ref{eq:Qapprox}). (2) Results from
 finite element numerical simulations using the thermodynamic approach of
 Section~\ref{section:thermodynamic_approach}. The numerical results are fit to
 the simple harmonic oscillator response given by Eq.~(\ref{eq:ud}) to determine the listed diagnostics.}
 \label{table:beam_fluid_sader}
\end{table}

\begin{figure}[h]
\begin{center}
\includegraphics[width=3.0in]{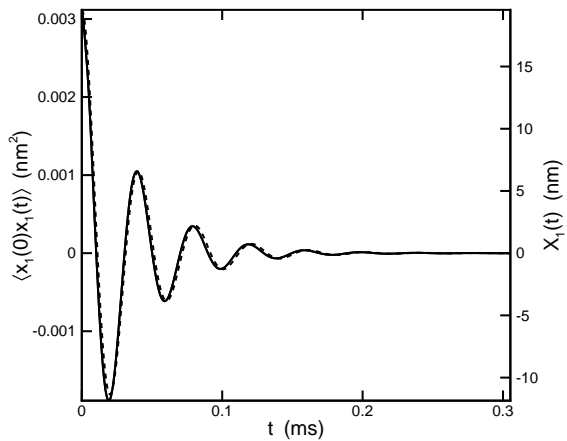}
\end{center}
\caption{The dynamics of a micron scale cantilever immersed in
water. (solid line) Deterministic finite element numerical
simulations using the thermodynamic approach. (dashed line) A
simple harmonic oscillator curve fit to the data. The left
ordinate yields the autocorrelations of the fluctuations in
cantilever displacement and the right ordinate yields the
deterministic cantilever dynamics in response to the removal of a
step force.} \label{fig:beam_fit}
\end{figure}

We now use the thermodynamic approach to calculate the stochastic
dynamics of the cantilever. As previously stated this can be
accomplished in a straightforward manner by determining the
deterministic response of the cantilever to the removal of a step
force applied to the tip. We do this using finite element
numerical simulations of the full three dimensional,
time-dependent, fluid-solid interaction problem (algorithm
discussed elsewhere~\cite{yang:1994,{cfdrc}}). The simulation is
initiated with the removal of a step force applied to the tip of
the cantilever and the deterministic dynamics of the beam $X_1(t)$
are shown by the solid line in Fig.~\ref{fig:beam_fit} using the
right ordinate axis.

In order to calculate commonly used diagnostics, such as
$\omega_f$ and $Q$, the deterministic cantilever deflection
$X_1(t)$ was fit to the deflections of an underdamped simple
harmonic oscillator (valid for $Q>1/2$) given by the equation of
motion,
\begin{equation}
 m_f \ddot{X_1}(t) + \gamma_f\dot{X_1}(t) + k X_1(t) = 0,
 \label{eq:sho_fit}
\end{equation}
where $X_1(0)=F_1/k$ and $\dot{X_1}(0)=0$. The solution is,
\begin{equation}
X_1(t) = \frac{F_1}{k} e^{-\omega_f t/2Q} \left( \cos(\omega' t) +
\frac{\omega_f}{2Q\omega'}\sin(\omega't) \right) \label{eq:ud}
\end{equation}
where,
\begin{equation}
\omega' = \omega_f \sqrt{1-\frac{1}{4Q^2}},
\end{equation}
and $\omega_f = \sqrt{k/m_f}$. Using a nonlinear least squares
curve fit algorithm~\cite{matlab} the numerical results are fit to
the model to yield the values of $Q$, $\omega_f$, $\gamma_f$, and
$m_f$ shown in Table~\ref{table:beam_fluid_sader}. The curve fit
is nearly indistinguishable from the numerical simulation results
for $X_1(t)$ and is shown by the dashed line in
Fig.~\ref{fig:beam_fit}.

\begin{figure}[h]
\begin{center}
\includegraphics[width=3.0in]{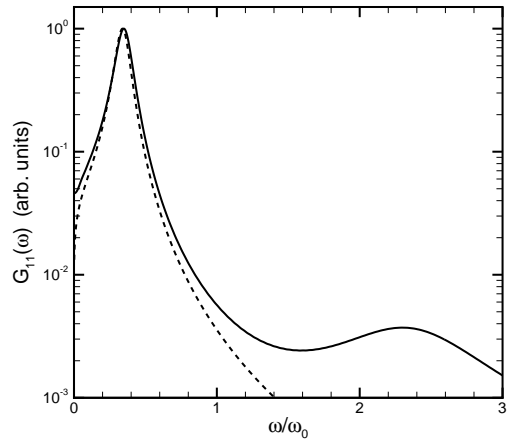}
\end{center}
\caption{The noise spectrum $G_{11}(\omega)$ for a micron scale
cantilever in water. (solid line) Deterministic finite element
numerical simulations using the thermodynamic approach. (dashed
line) Approximate analytical theory for the fundamental mode
only.} \label{fig:beam_noise}
\end{figure}

Using $X_1(t)$ in Eq.~(\ref{eq:auto}) yields the autocorrelation
of the equilibrium deflections, $\left< x_1(0) x_1(t) \right>$,
which are shown in Fig.~\ref{fig:beam_fit} using the left ordinate
axis. The noise spectrum is given by Eq.~(\ref{eq:noise1}) and is
shown in comparison with predictions of the infinite-beam
approximation given by Eq.~(\ref{eq:gxf_noise}) in
Fig.~\ref{fig:beam_noise}. The numerical simulations contain all
of the oscillation modes as indicated by the presence of the
second mode in the figure. In the infinite beam approximation the
different modes behave independently and the calculation could be
extended to include higher modes if desired. In
Table~\ref{table:beam_fluid_sader} the analytical results of
Section~\ref{section:analytical_theory} are compared with results
using the full thermodynamic approach and deterministic finite
element numerical simulations. In general, from the table and the
figures, we see that the differences between the simple model and
the full calculations are small for this shape cantilever. The
experimentally measured value of the resonant frequency in water
is $\omega_f/\omega_0=0.36$~\cite{chon:2000} which agrees well
with both the analytical results and the numerical predictions
using the thermodynamic approach. However,
Table~\ref{table:beam_fluid_sader} suggests that the analytical
results based upon the oscillating infinite cylinder model under
predict the amount of damping $\gamma_f$. The noise spectra show
differences that are small, but may be significant for
quantitative applications such as calibration.

In all of the deterministic finite element numerical simulations
performed we were careful to ensure that the bounding no-slip
surfaces of the computational domain did not affect the results.
Comparisons with numerical simulations performed with larger
domains did not significantly alter the results.

\section{An array of nanoscale cantilevers}
\label{section:nanoscale_cantileves} As the cantilever dimensions
become smaller the effective cantilever spring constant decreases
while the resonant frequency increases. This favorable combination
potentially provides access to the biologically important
parameter regime characterized by 10's of piconewtons with
microsecond scale time resolution, a range that is difficult to
reach using other methods. This has led to the development of
nanoscale
cantilevers~\cite{arlett:2005,roukes:2000:1,craighead:2000}.

\begin{figure}[tbh]
\begin{center}
\includegraphics[width=2.75in]{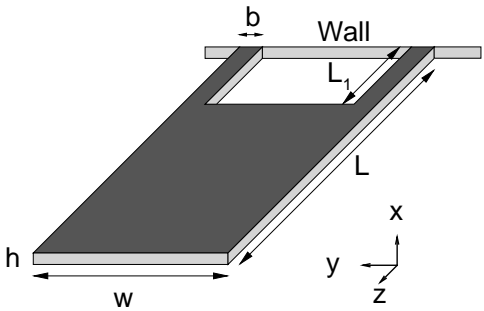}
\end{center}
\caption{Schematic of an experimentally motivated nanoscale
cantilever geometry of length $L$, width $w$, and height $h$.}
\label{fig:paddle_config}
\end{figure}

In what follows we quantify the stochastic dynamics of two
adjacent nanoscale cantilevers immersed in water. A schematic of
the nanoscale cantilever under consideration here is shown in
Fig.~\ref{fig:paddle_config}~\cite{arlett:2005,roukes:2000:1}.
This is an experimentally motivated cantilever whose paddle shaped
geometry decreases the effective spring constant while localizing
the strain for piezoresistive measurement. The configuration of
the cantilever array is shown in Fig.~\ref{fig:adjacent} with two
opposing cantilevers separated by a distance $s$. This
configuration was chosen because of its experimental accessibility
as well as its potential use in single molecule measurements with
the tethering of a target biomolecule between the two cantilevers.
In the work presented here we build a baseline understanding of
the cantilever dynamics in the absence of target biomolecules.
Further consideration of the dynamics caused by a tethered
biomolecule is beyond the scope of the present paper. The
dimensions of the nanoscale cantilevers investigated here are
summarized in Table~\ref{table:paddle-geometry}.

As discussed previously, the physical properties describing the
dynamics of the cantilever in vacuum are important parameters in
determining the stochastic dynamics of the cantilevers. These
properties are summarized in Table~\ref{table:paddle_vacuum} where
the values have been determined from finite element simulations of
the elastic cantilever structure in the absence of the surrounding
fluid. It is important to emphasize that although these
calculations can sometimes be performed analytically, this is not
the case for many geometries of interest. However, even for
cantilevers with very complex geometries, the vacuum based finite
element calculations are straight forward.
\begin{table}
[h]
\begin{center}
\begin{tabular}
[c]{l@{\hspace{0.5cm}}l@{\hspace{0.5cm}}l@{\hspace{0.5cm}}l@{\hspace{0.5cm}}l@{\hspace{0.5cm}}l}
  $L$   & $w$    & $h$  & $L_1$     & $b$ \\ \hline \hline \\
   3$\mu$m & 100nm & 30nm & $0.6\mu$m & 33nm \\
\end{tabular}
\end{center}
 \caption{Geometry of the nanoscale cantilever (see Fig.~\ref{fig:paddle_config}).
The nanoscale cantilever is composed of silicon ($E = 1.25 \times
10^{11}$N/m$^2$, $\rho_c = 2330$kg/m$^3$) and the fluid is water
($\rho_f = 997$kg/m$^3$, $\eta = 8.67 \times 10^{-4}$kg/ms). The
magnitude of the step force applied to the tip of the cantilever
is $F_0 = 0.75$pN.}
 \label{table:paddle-geometry}
\end{table}
\begin{table}
[h]
\begin{center}
\begin{tabular}
[c]{l@{\hspace{0.5cm}}l@{\hspace{0.5cm}}l@{\hspace{0.5cm}}l@{\hspace{0.5cm}}l@{\hspace{0.5cm}}l@{\hspace{0.5cm}}l}
  $k$       & $\omega_0$   & $R_0$ & $T_0$ \\ \hline \hline \\
   8.7 mN/m & $37.46 \times 10^6$ rads/s & 0.11  & 1.12 \\
\end{tabular}
\end{center}
 \caption{The cantilever spring constant $k$, resonant frequency in vacuum $\omega_0$,
  the Strouhal number $R_0$, and the mass loading factor $T_0$. $\omega_0$ and $k$
were determined
  from numerical simulations of the nanoscale cantilever in vacuum.}
 \label{table:paddle_vacuum}
\end{table}

The stochastic dynamics of the cantilever array are determined
using the thermodynamic approach discussed in
Section~\ref{section:thermodynamic_approach} with deterministic
finite element numerical simulations. A series of numerical
simulations have been performed to determine the variation in the
cantilever dynamics as a function of the separation distance
between the two cantilevers in the array. In particular we have
explored $s/h = 1,2,3,4,5$ where $s$ is the cantilever separation
and $h=30$nm is the cantilever thickness. The numerical
simulations are initiated by the removal of a step force applied
to the tip of one of the cantilevers (in Fig.~\ref{fig:adjacent}
this is the cantilever on the left) while the other cantilever is
initially undeflected and at mechanical equilibrium.

\begin{figure}[h]
\begin{center}
\includegraphics[width=3.0in]{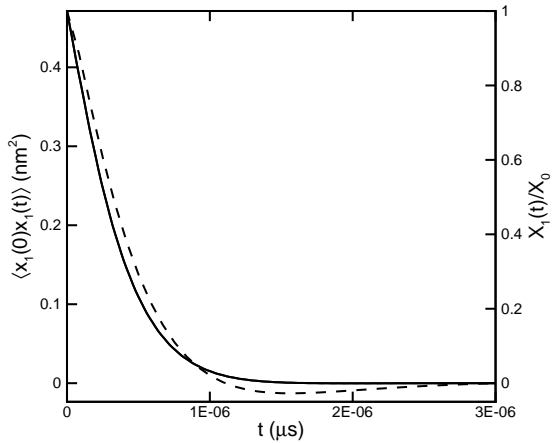}
\end{center}
\caption{Comparison between the model theory and the full
numerical simulation for the autocorrelation of equilibrium
fluctuations (right ordinate axis) and the deterministic
cantilever deflection relative to the initial deflection of $X_0 =
X_1(t=0)$ (left ordinate axis) for a single nanoscale cantilever
in fluid: (solid line) results from finite element numerical
simulations using the thermodynamic approach; (dashed line)
results based upon the approximate analytical theory of
Section~\ref{section:analytical_theory}.}
\label{fig:one_paddle_auto1}
\end{figure}

The deterministic dynamics of the nanoscale cantilever for which
the step force has been removed, $X_1(t)$, is shown by the solid
line in Fig.~\ref{fig:one_paddle_auto1} using the right ordinate
axis. The deflection $X_1$(t) was not significantly affected by
the presence of the second cantilever for the separations
considered here. As with the micron scale cantilever, we again fit
the result for $X_1(t)$ from the finite element numerical
simulation to the solution of the simple harmonic oscillator
equation given by Eq.~(\ref{eq:sho_fit}). For the nanoscale
cantilever the dynamics are overdamped ($Q<1/2)$ and the
deterministic deflection predicted by the simple harmonic
oscillator approximation is,
\begin{equation}
X_1(t) = \frac{F_1}{k} \left(
\frac{\lambda_2}{\lambda_2-\lambda_1}e^{\lambda_1t} +
\frac{\lambda_1}{\lambda_1-\lambda_2}e^{\lambda_2t} \right)
\label{eq:od}
\end{equation}
where,
\begin{equation}
\lambda_{1,2} = \omega_f \left( -\frac{1}{2Q} \pm
\sqrt{\frac{1}{4Q^2}- 1}\right).
\end{equation}
The agreement with the curve fit is very good and $\omega_f$ and
$Q$ are shown in Table~\ref{table:paddle_fluid}. Also shown in
Table~\ref{table:paddle_fluid} are the theoretical predictions
based upon the infinite cylinder approximation. For the analytical
calculations of the damping $\gamma_f$ from Eq.~(\ref{eq:damping})
the fundamental mode of the nanoscale cantilever has been modelled
as that of a hinge where all of the bending occurs in the short
legs near the base. This is guided by numerical calculations
showing the strain localized in the short legs. In this case we
find the mass factor is $\alpha=1/3$. (Note that
$\alpha$ is not needed to determine the other quantities in
Table~\ref{table:paddle_fluid}). It is clear from
Table~\ref{table:paddle_fluid} that the infinite cylinder
approximation is no longer valid. The analytical prediction for
the added mass is an order of magnitude too large. This suggests
that three-dimensional flow effects become significant for the
shape of cantilever under consideration here. For example there
would be flow around the tip of the cantilever, as well as the
flow through the open region near the base of the cantilever. In
addition, the predicted frequency shift is an order of magnitude
too small.
\begin{table}
[h]
\begin{center}
\begin{tabular}[c]{l@{\hspace{0.5cm}}l@{\hspace{0.5cm}}l@{\hspace{0.5cm}}l@{\hspace{0.5cm}}l@{\hspace{0.5cm}}l@{\hspace{0.5cm}}l}
  & $Q$ & $m_f/m_e$ & $\omega_f/\omega_0$ & $\gamma_f$ (kg/s) & $R_f$  \\
  \hline \hline \\
  (1) & 0.265 & 106.3 & 0.032 & $2.41 \times 10^{-9}$ & 0.0036   \\
  (2) & 0.327 & 18.16 & 0.236 & $1.01 \times 10^{-9}$ & 0.026 \\
\end{tabular}
\end{center}
 \caption{The stochastic dynamics of a nanoscale cantilever in fluid. (1) Results
 based on the analytical predictions of Section~\ref{section:analytical_theory}. The quality $Q$
 is calculated from the approximation given by Eq.~(\ref{eq:Qapprox}). (2) Results from
 finite element numerical simulations using the thermodynamic approach of
 Section~\ref{section:thermodynamic_approach}. The numerical results are fit to the simple
 harmonic oscillator response given by Eq.~(\ref{eq:od}) to determine the listed diagnostics.}
 \label{table:paddle_fluid}
\end{table}

\begin{figure}[h]
\begin{center}
\includegraphics[width=3.0in]{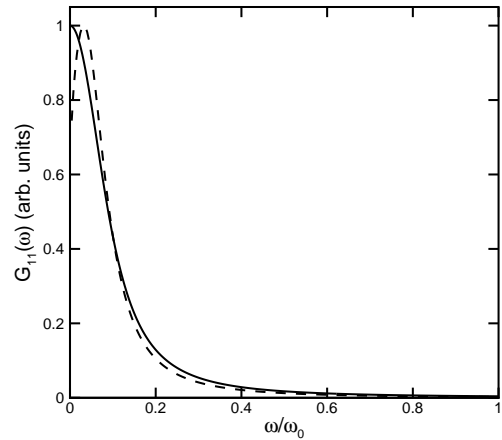}
\end{center}
\caption{The noise spectrum $G_{11}(\omega)$ for a single
nanoscale cantilever immersed in water. (solid line) Results from
deterministic finite element simulations using the thermodynamic
approach described in
Section~\ref{section:thermodynamic_approach}. (dashed line)
Results from approximate analytical theory described in
Section~\ref{section:analytical_theory}.}
\label{fig:one_paddle_noise_comparison}
\end{figure}

Inserting $X_1(t)$ into Eq.~(\ref{eq:auto}) yields the
autocorrelation of the equilibrium fluctuations in cantilever
displacement, $\left< x_1(0) x_1(t) \right>$. These are also shown
by the solid line in Fig.~\ref{fig:one_paddle_auto1} using the
ordinate axis on the left. The noise spectrum, $G_{11}(\omega)$,
is found by inserting the autocorrelation in Eq.~(\ref{eq:noise1})
and the result is shown by the solid line in
Fig.~\ref{fig:one_paddle_noise_comparison}.
Fig.~\ref{fig:one_paddle_noise_comparison} also presents a
comparison of the noise spectrum $G_{11}(\omega)$ given from the
theoretical predictions of Eq.~(\ref{eq:gxf_noise}) and the
results from the numerical simulations using the thermodynamic
approach. The noise spectra are different for small frequencies,
in particular the model result has a peak at finite frequency not
seen in the full calculation. The inverse cosine transform of the
noise spectrum from the infinite-cylinder model yields the
deflection of the cantilever as a function of time shown by the
dashed line in Fig.~\ref{fig:one_paddle_auto1} using the right
ordinate axis. The model prediction for the cantilever deflection
exhibits negative values that are not seen in the simulations
suggesting that the response of the overdamped cantilever is not
precisely that of a simple harmonic oscillator.

\begin{figure}[h]
\begin{center}
\includegraphics[width=3.0in]{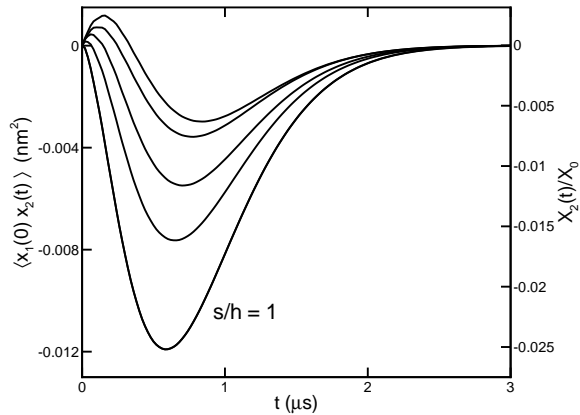}
\end{center}
\caption{The cross-correlations in equilibrium fluctuations of
cantilever deflections for two adjacent nanoscale cantilevers
separated by distances $s/h=1,2,3,4,5$ for $h=30$nm found using
deterministic finite element numerical simulations with the
thermodynamic approach discussed in
Section~\ref{section:thermodynamic_approach}. The curve for
$s/h=1$ is labelled and the others follow in sequential order. The
right ordinate axis illustrates the deterministic cantilever
deflection as a function of time $X_2(t)$ scaled by the deflection
of the adjacent cantilever, $X_0=X_1(t=0$).}
\label{fig:two_paddles_cross_all}
\end{figure}

We now turn to the deterministic deflection of the second cantilever, $X_2(t)$,
after removing the force on the first cantilever, evaluated using the full
fluid-elasticity simulations. This is shown in
Fig.~\ref{fig:two_paddles_cross_all} using the right ordinate
axis. For close separations $s/h \lesssim 1$ the second cantilever
exhibits negative deflections for all time. However, as the
cantilever separation increases the second cantilever initially
exhibits a positive deflection followed by a negative deflection.
The flow field around an oscillating object will have both
potential and non-potential components. For an incompressible
fluid the potential component is instantaneous whereas the
non-potential component diffuses with a diffusion coefficient
given by the kinematic viscosity $\nu$. Although the deterministic
flow field around simple oscillating objects is well known the
fluid coupled motion of multiple elastic objects is not well
understood. The fluid dynamics resulting from the motion of two
adjacent cantilevers will be discussed in a forthcoming
article~\cite{clark:2006}.

Using Eq.~(\ref{eq:cross}) the stochastic correlations between the
two cantilevers can be determined from $X_2(t)$. The
cross-correlations of the equilibrium displacement fluctuations
$\left<x_1(0)x_2(t)\right>$ are shown in
Fig.~\ref{fig:two_paddles_cross_all} using the ordinate axis on
the left. For small cantilever separations $s/h \lesssim 1$ the
cross-correlations are anticorrelated for all time. This is in
agreement with experimental measurements of the fluid correlations
of two closely spaced micron scale beads in
water~\cite{meiners:1999,meiners:2000}. However, as the cantilever
separation is increased the cross-correlations changes: for short
times the cross-correlations are positively correlated whereas for
larger times they are anti-correlated.

\begin{figure}[h]
\begin{center}
\includegraphics[width=3.0in]{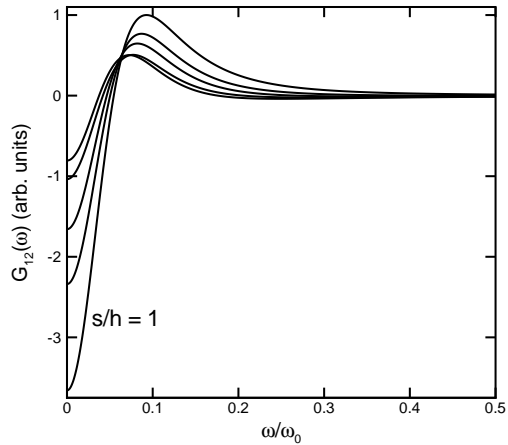}
\end{center}
\caption{The noise spectrum $G_{12}(\omega)$ for two nanoscale
cantilevers separated by distance $s/h=1,2,3,4,5$ for $h=30$nm
found using deterministic finite element numerical simulations
with the thermodynamic approach discussed in
Section~\ref{section:thermodynamic_approach}. The curve for
$s/h=1$ is labelled and the others follow in sequential order.}
\label{fig:two_paddles_noise_all}
\end{figure}

The noise spectra $G_{12}(\omega)$ are found from
$\left<x_1(0)x_2(t)\right>$ by using Eq.~(\ref{eq:noise2}).
Figure~\ref{fig:two_paddles_noise_all} shows the variation in the
noise spectra as a function of cantilever separation. The noise
spectra contain both positive and negative values and for each
cantilever separation there is a frequency at which correlated
noise vanishes, $G_{12}(\omega)=0$. This null point could be
exploited by a measurement scheme to minimize the correlated
noise.
\begin{figure}[h]
\begin{center}
\includegraphics[width=2.75in]{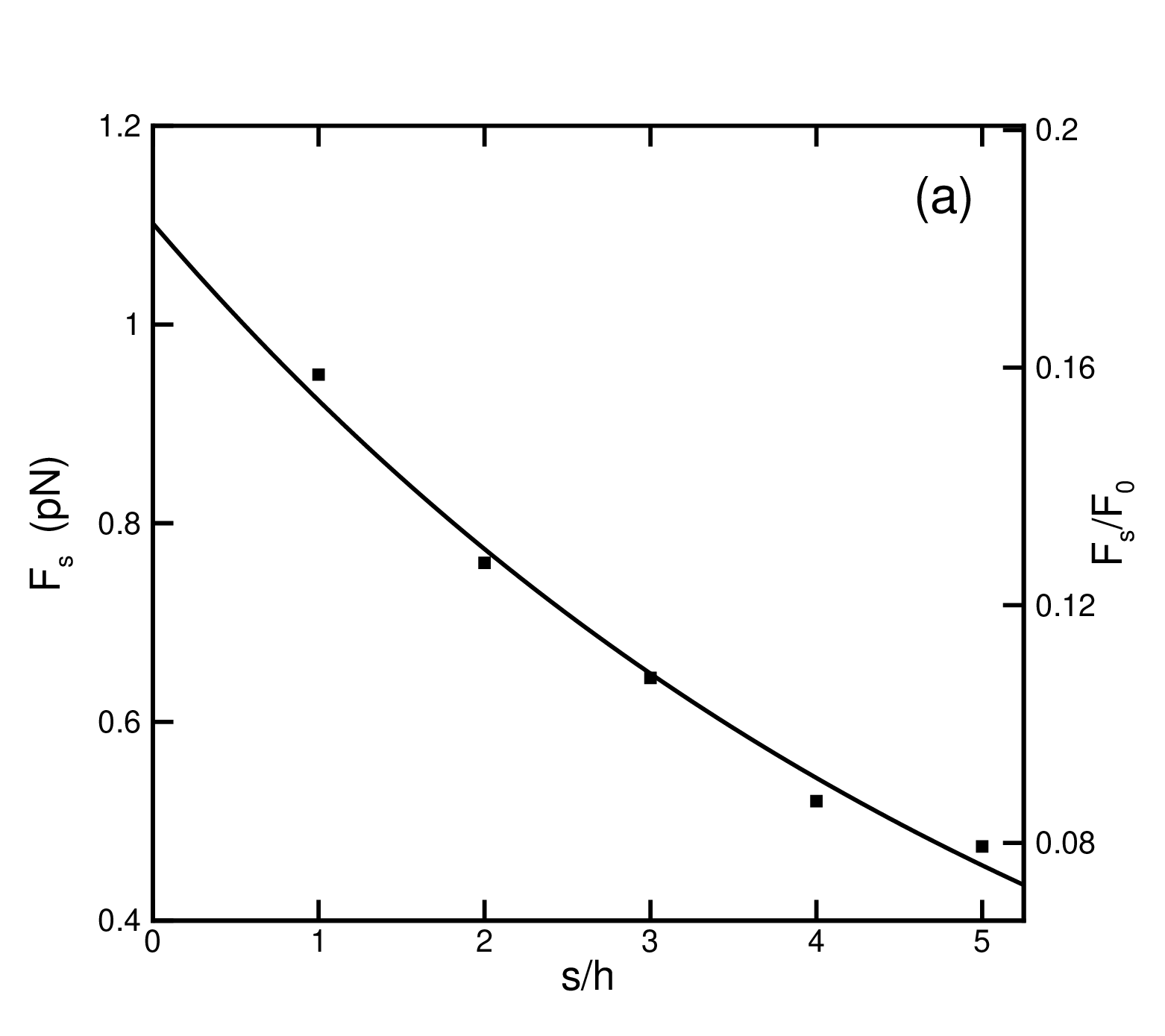}
\includegraphics[width=2.75in]{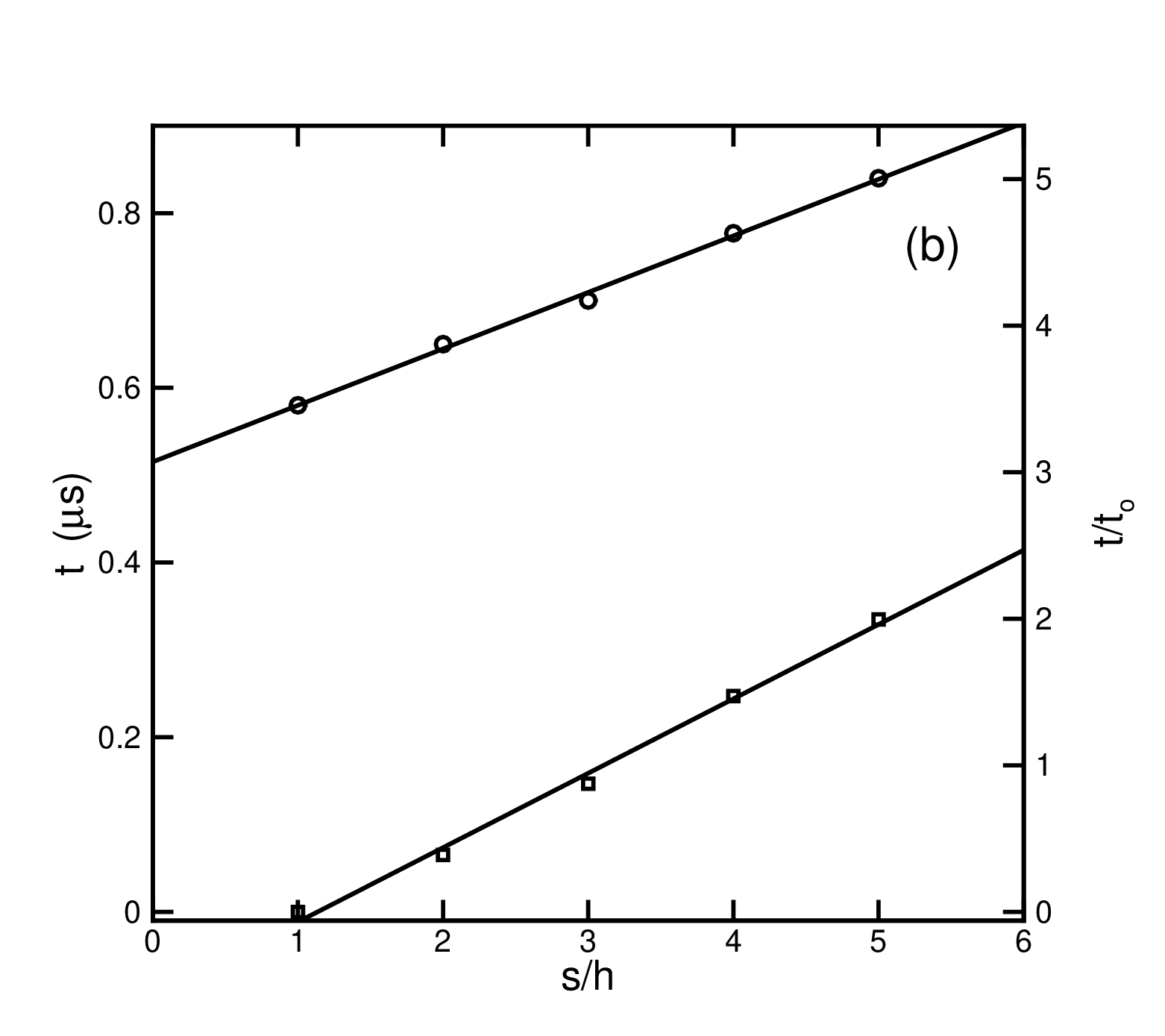}
\end{center}
\caption{The force sensitivity and time scales characterizing two
adjacent nanoscale cantilevers found using deterministic finite
element numerical simulations with the thermodynamic approach
discussed in Section~\ref{section:thermodynamic_approach}.
Panel~(a), the maximum magnitude of the force induced by
fluid-coupled correlations in cantilever displacement $F_s$. The
data is empirically fit with an exponential given by $F_s = 1.12
\exp(-0.18s/h)$ (in units of pN). The magnitude of the force felt
by a single cantilever due to Brownian noise, $F_0=6.0$ pN, is
nearly 6 times larger than the correlated noise at close
separations ($s/h \lesssim 1$). The right ordinate axis scales
$F_s$ by $F_0$. Panel~(b), characteristic time scales, the circles
represent the time at which the maximum value of the cross
correlation occurs, the squares represent the time at which the
fluid induced correlations vanish. The ordinate axis on the right
hand side illustrates the time scales when scaled by the period of
oscillation of the cantilever in vacuum $t_0=(\omega_0/2
\pi)^{-1}$. The solid lines represent a linear curve fit to the
data.} \label{fig:force_and_time}
\end{figure}

From these results it is possible to characterize the force
sensitivity and time resolution of a correlation measurement
technique using an array of closely spaced nanoscale cantilevers.
An estimate of the force sensitivity can be found using the auto
and cross correlation functions as,
\begin{eqnarray}
  \label{fsens}
  F_0 &=& k ||\langle x_1(0) x_1(t) \rangle||_{max}^{1/2}, \\
  F_s &=& k ||\langle x_1(0) x_2(t) \rangle||_{max}^{1/2}.
  \label{eq:force_scales}
\end{eqnarray}
$F_0$ represents the approximate magnitude of the stochastic
Brownian force acting on a single cantilever and $F_s$ the force
induced by fluid correlations between two cantilevers. For the
case of a single nanoscale cantilever the maximum value occurs at
$t=0$ and, using the equipartition theorem this yields a value of
$F_0 = \sqrt{k_B T k} = 6.0$pN.

In a cross-correlation measurement between two cantilevers the
Brownian noise felt by the two individual cantilevers is
uncorrelated and does not contribute. This leaves only the
correlations due to the viscous fluidic coupling and $F_{s}$
represents the approximate magnitude of the force due to this
hydrodynamic coupling. For example, from
Fig.~\ref{fig:two_paddles_cross_all} for the case with the closest
separation $s/h=1$, the magnitude of the maximum value of the
cross-correlation is $\langle x_1(0) x_2(t) \rangle = 1.19 \times
10^{-2} $nm$^2$ which yields a force sensitivity $F_{s} = 0.95$pN.
Therefore $F_s/F_0 \approx 1/6$ at $s/h=1$, indicating a 6-fold
reduction in thermal noise over a single cantilever measurement.

The variation in $F_s$ as a function of cantilever separation is
shown in Fig.~\ref{fig:force_and_time}(a). The data is fit with an
exponential curve, given by $F_s = 1.12 \exp(-0.18s/h)$ (in units
of pN), shown in Fig.~\ref{fig:force_and_time}(a) as the solid
line. For very small cantilever separations $s/h \lesssim 1$ the
noise reduction possible through the use of a correlated
measurement technique is approximately 6-fold. As expected, the
correlated noise decreases as the cantilever separation is
increased: for a separation of $s/h=5$ there is a 12-fold noise
reduction. The ordinate axis on the right hand side of
Fig.~\ref{fig:force_and_time}(a) shows $F_s/F_0$ to show this
performance improvement directly.

The characteristic time scales of the fluid correlated motion as a
function of cantilever separation are illustrated in
Fig.~\ref{fig:force_and_time}(b). In
Fig.~\ref{fig:force_and_time}(b) circles represent the time at
which the maximum value in the cross correlation occur and squares
represent the time at which the fluid induced correlations vanish
(i.e., the zero-crossing in Fig.~\ref{fig:two_paddles_cross_all}).
As expected from the finite rate at which momentum diffuses away
from an oscillating cantilever, both of these time scales increase
with cantilever separation. Using the ordinate axis on the right
in Fig.~\ref{fig:force_and_time}(b) the ratio of these time scales
to the period of a single oscillation for the cantilever in vacuum
$t_0 = f_0^{-1}$ are given.

\section{Conclusions}
The stochastic dynamics of micron and nanoscale cantilevers can be
quantified for the precise conditions of experiment using
straightforward deterministic calculations coupled with the
fluctuation-dissipation theorem. This can be done for complex
cantilever geometries including wall effects, and the stochastic
correlated behavior of closely spaced cantilevers for which
theoretical predictions currently are not available. Cantilever
geometries that are long and thin are well described by
approximate analytical theory based upon the oscillation of an
infinite cylinder in fluid. We have used the fluctuation-dissipation method to
correct the previous results of this theory. For the short and wide
nanocantilevers currently being proposed the infinite-cylinder approximation is no
longer appropriate. The thermodynamic approach using the
fluctuation-dissipation theorem provides an important means of
developing a physical understanding of the stochastic dynamics of
closely spaced micron and nanoscale objects that will be important
as micron and nanotechnology progress and a quantitative design
tool for experiment.

This research has been partially supported by DARPA/MTO Simbiosys
under grant F49620-02-1-0085 and an ASPIRES grant from Virginia
Tech. This work has been carried out in collaboration with the
Caltech BioNEMS effort (M. L. Roukes, PI) and we gratefully
acknowledge extensive interactions with this team.

\end{document}